\documentclass[twocolumn,superscriptaddress,amsmath,amssymb,floatfix,prb,showpacs]{revtex4}
\usepackage{amsmath,amssymb,natbib,bm,graphicx,url}
\usepackage[ansinew]{inputenc}

\newcommand{\be}{\begin{equation}}
\newcommand{\ee}{\end{equation}}
\newcommand{\bes}{\begin{equation}\begin{split}}
\newcommand{\ees}{\end{split}\end{equation}}

\newcommand{\vc}[1]{\mathbf{#1}}

\newcommand{\abs}[1]{\left|#1\right|}

\newcommand{\ket}[1]{\left|\, #1 \, \right\rangle}

\newcommand{\boket}[3]{\left\langle\, #1 \,|\, #2 \,|\, #3 \,\right\rangle}

\DeclareMathOperator{\ii}{i}

\DeclareMathOperator{\rre}{Re}
\DeclareMathOperator{\iim}{Im}
\DeclareMathOperator{\sgn}{sgn}

\begin{document}
\title{Thermopower of Single-Molecule Devices}
\author{Jens Koch}
\author{Felix \surname{von Oppen}}
\affiliation{Institut f\"ur Theoretische Physik, Freie Universit\"at Berlin, Arnimallee 14, 14195 Berlin, Germany}
\author{Yuval Oreg}
\author{Eran Sela}
\affiliation{Department of Condensed Matter Physics, Weizmann Institute for Science,  Rehovot 76100,
Israel}
\date{November 14, 2004}
\begin{abstract}
We investigate the thermopower of single molecules weakly coupled to 
metallic leads. We model the molecule in terms of the relevant 
electronic orbitals coupled to phonons corresponding to
both internal vibrations and to oscillations of the molecule as a whole. 
The thermopower is
computed by means of rate equations including both sequential-tunneling 
and cotunneling processes. Under certain conditions, the thermopower 
allows one to access the electronic
and phononic excitation spectrum of the molecule in a linear-response 
measurement.
In particular, we find that the phonon features are more pronounced for 
weak lead-molecule
coupling. This way of measuring
the excitation spectrum is less invasive than the more conventional 
current-voltage
characteristic, which,  by contrast, probes the system far from equilibrium.
\end{abstract}
\maketitle
\section{Introduction}
The physics of electronic transport through single molecules has gained much interest in recent years, both experimentally and theoretically. Experiments with molecules ranging from $\mathrm{H_2}$ \cite{ruitenbeek} to DNA \cite{porath} have shown various interesting effects occurring in two-terminal \cite{ruitenbeek,porath,weber,chen} and three-terminal \cite{dekker,park,park2,pasupathy,natelson} molecular devices. The spectrum includes Coulomb blockade,\cite{park2,natelson} negative differential resistance (NDR),\cite{chen} phonon influences,\cite{park,pasupathy,natelson} and the Kondo effect.\cite{park2,natelson} The ultimate goal of these efforts is to realize the vision of molecular electronics.\cite{reedt} While the possibilities of concrete applications remain to be seen, the question of transport in the molecular regime is of fundamental physical interest.

Recently, there has been considerable theoretical ef\-fort to calculate cur\-rent-vol\-tage ($IV$) characteris\-tics of single-molecule devices, and effects such as  NDR,\cite{schoeller1,schoeller2,schoeller3} influences of phonons and dissipation,\cite{aleiner,varma,schoeller3,flensb1,flensb2} and contact-geometry effects \cite{ratner2,diventra} have been studied intensely. Roughly speaking, present approaches separate into two main directions:
(1) Work based on a detailed modeling of the molecule and the contact region via density functional theory (DFT).\cite{ratner2,diventra} 
For a recent discussion about the validity of equilibrium DFT for transport through single molecules cf.~Ref.~\onlinecite{evers}.
(2) Approaches based on a parametric modeling of relevant molecular levels.\cite{schoeller1,schoeller2,schoeller3,aleiner,varma,flensb1,flensb2} This type of approach has the advantage that it enables the investigation of additional degrees of freedom including, e.g., mechanical, and magnetic degrees of freedom leading to phonon and spin dynamics. The present paper follows this second approach.

In this paper, we investigate the thermopower of single-molecule devices. The thermopower is defined as the ratio of voltage $V$ and an applied temperature difference $\Delta T$  under the condition that the current vanishes:
\be\label{thpdef}
S=-\lim_{\Delta T\rightarrow 0} \frac{V}{\Delta T}\bigg|_{I=0}.
\ee
For quantum dots the thermopower for pure sequential tunneling has been investigated theoretically by Beenakker and Staring\cite{beenakker1} and experimentally by Staring \emph{et al}.\cite{beenakker2} The cotunneling regime\cite{averin2} and the crossover have been studied by Turek and Matveev.\cite{turek} In the case of a quantum dot strongly coupled to one lead, the thermopower of quantum dots has been investigated by Matveev and Andreev.\cite{andreev}  

Here, we extend these considerations to transport through single molecules, where experimental work\cite{park,pasupathy,natelson} indicates that phonons may play an important role. In our model, we consider transport through relevant electronic orbitals and incorporate Coulomb interaction by a Hubbard-like term. These electronic orbitals are coupled to both oscillations of the molecule relative to the leads and internal vibrations. The coupling of the molecule to the leads is represented by a tunneling Hamiltonian. Employing the rate-equation approach valid for weak molecule-lead coupling, we compute the thermopower as a function of gate voltage, temperature, and electron-phonon coupling.

We find that the thermopower contains information on the electronic and phononic excitations of the molecule. This way of measuring the molecular excitations in linear response (cf.~the $\Delta T\rightarrow 0$ limit in Eq.~\eqref{thpdef}) may have advantages over the more conventional $IV$ characteristic. The latter necessarily involves nonequilibrium effects, which are difficult to interpret. Moreover, a large applied voltage may affect symmetry and structure of the molecule itself.

In leading order perturbation theory for the molecule-lead coupling, electrons tunnel from a lead onto the molecule or vice versa (sequential-tunneling contributions). For pure sequential tunneling we find that the thermopower as a function of gate voltage develops a sawtooth behavior in the low temperature limit with steps due to electronic and phononic excitations.  Step sizes and their dependence on the electron-phonon coupling strength are analyzed. 

Moreover, we find that in a wide range of parameters so-called cotunneling contributions from next-leading order perturbation theory for the molecule-lead coupling are important.\cite{footnote} In this case, the electron only virtually occupies molecular levels. We investigate the cotunneling contributions and the full crossover between the sequential-tunneling and the cotunneling regimes. We find that elastic cotunneling does not show any significant phonon structure, and discuss under which conditions sequential-tunneling phonon features are retained in the total thermopower. 

The outline of this paper is as follows: Section \ref{sec:model} introduces our model for a single-molecule device with mechanical degrees of freedom. In section \ref{sec:rate} we review the rate equations approach and address the issue of regularization of the cotunneling contributions. Our calculations for the thermopower are described in section \ref{sec:thp} and the results are presented in section \ref{sec:res}. We summarize our findings in section \ref{sec:sum}. Some calculational details including the results of the cotunneling regularization are relegated to appendices.

\section{Model\label{sec:model}}
We consider a three-terminal single-molecule device, consisting of a molecule weakly coupled to two metallic leads serving as source and drain electrode, respectively. The third electrode only influences the molecule by electrostatic interaction and serves as a gate electrode. In order to measure the thermopower of the device, it is necessary to control the electrostatic potentials and temperatures of the source and drain electrodes individually.

The model we apply in order to investigate the thermopower has previously been used in analyses of $IV$ characteristics, see, e.g., Refs.~\onlinecite{aleiner,varma}. The Hamiltonian can be divided into a part describing the electronic and phononic features of the molecule, a part modeling the leads, and a tunneling term that couples molecule and leads, $H= H_\text{mol} + H_\text{leads} + H_\text{mix}$, where
\begin{align}
H_\text{mol}= &(\varepsilon-eV_g) n_d + \frac{U}{2} n_d(n_d-1) \nonumber\\
&+ \lambda \hbar\omega_\text{vib} (b^\dag + b)n_d + \hbar\omega_\text{vib}(b^\dag b+1/2)  \label{Hmol}\\
&+ \frac{p_z^2}{2M} +\frac{1}{2}M\omega_\text{osc}^2 z^2\;, \nonumber\\
H_\text{leads}= &\sum_{a=L,R}\sum_{\vc{p},\sigma} \epsilon_\vc{p} c^\dag_{a\vc{p}\sigma}c_{a\vc{p}\sigma}\;,\\  
H_\text{mix}= &\sum_{a=L,R}\sum_{\vc{p},\, \sigma} \left( t_a(z) c^\dag_{a\vc{p}\sigma} d_\sigma + \text{h.c.}\right).\label{Hmix}
\end{align}
In the following, the common Fermi energy of the leads at vanishing bias voltage is chosen as the zero point of energy. For simplicity, we assume that only a single spin-degenerate orbital of the molecule with one-particle energy $\varepsilon$  contributes to the current. (A generalization towards more orbitals is not difficult.) For double occupancy of the molecule, Coulomb blockade is taken into account via the charging energy $U$. 
The operator $d_\sigma$ ($d_\sigma^\dag$) annihilates (creates) an electron with spin projection $\sigma$ on the molecule, $n_d=\sum_\sigma d_\sigma^\dag d_\sigma$ denotes the molecule occupation-number operator. 
The whole system of molecular levels can be shifted by means of applying a gate voltage $V_g$.

The leads are described as a non-interacting Fermi gas of electrons with a constant density of states. Here, $c_{a\vc{p}\sigma}$ ($c_{a\vc{p}\sigma}^\dag$) annihilates (creates) an electron in lead $a$ ($a=L,R$) with momentum $\vc{p}$ and spin projection $\sigma$. The potential $V_a$ and temperature $T_a$ of the left and right lead are taken into account through the probability distributions for state occupation in the leads. It is assumed that relaxation in the leads is sufficiently fast so that at any time these distributions have the form of Fermi functions:
\be
f_a(E) = \left( \exp[ (E+eV_a)/k_BT_a ] +1 \right)^{-1}.
\ee

We distinguish two types of phonons, which we term vibrations and oscillations: Vibrations are internal phonon modes of the molecule, for which the center of mass (CM) of the molecule is at rest, while oscillations involve movement of the molecule as a whole. Vibrational phonons are annihilated (created) by $b$ ($b^\dag$). For oscillations we use the momentum and position operators $p_z$ and $z$ of the CM displacement. In the case of phy\-sisorbed molecules \footnote{
Physisorption refers to weak bonding of a molecule to a surface by van-der-Waals forces or hydrogen bridge bonds, as opposed to chemisorption, which terms the case of strong covalent bonds between molecule and surface.} the coupling to the leads is weak, so that these two phonon types typically involve different energy scales: Vibrations, which are associated with strong intra-molecular bonds, will have considerably higher energies than oscillations. The two phonon types also differ in the nature of coupling: Vibrations directly couple to the electric charge on the molecule, described by the term $\sim n_d(b^\dag + b)$, whereas the coupling for oscillations occurs through displacement-dependent tunneling matrix elements $t_{L,R}(z)$.

(a) \emph{Oscillations.}---Since $t$ arises due to tunneling processes between the leads and the molecule, we assume an exponential fall-off of $t$ with increasing distance between lead and molecule. For a symmetric molecule of length $2l$ between two leads with a separation distance $2d$, this yields 
\be\label{telem} t_{L,R}(z)=t_0\exp[-(d-l\pm z)/z_0].\ee
The parameter $z_0$ fixes the length scale of the exponential fall-off of the electronic wave functions outside the leads and the molecule. 
 
(b) \emph{Vibrations.}---For the vibrational electron-phonon coupling there exists a procedure which eliminates the coupling term by a canonical transformation of the Hamiltonian.\cite{mahan, aleiner} 
This yields a renormalization of the parameters $\varepsilon$ and $U$, and of the lead-molecule coupling $t(z)\rightarrow t(z) \exp[-\lambda(b^\dagger-b)]$. 
Henceforth, for the sake of notational simplicity, we will refer to the renormalized parameters as $\varepsilon$ and $U$. (Alternatively, one can diagonalize the Hamiltonian for each occupation number $n_d$ and calculate Franck-Condon matrix elements.)

In the following, we restrict ourselves to considering one phonon type at a time. Whenever the specific phonon type is irrelevant we will skip the subscripts ``vib" and ``osc".

\section{Rate equations and transition rates\label{sec:rate}}
We consider the weak-coupling regime for the mo\-le\-cule-lead coupling. In this case the energy broadening $\gamma$ of molecular levels due to $H_\text{mix}$ is the smallest energy in the problem. In particular, we assume $\gamma\ll k_BT,\hbar\omega$, which allows for a perturbative treatment for $H_\text{mix}$. 

In the absence of coupling to the leads, the eigenstates of the molecule can be written as $\ket{n\sigma,q}$, where $n$ denotes the number of additional electrons on the molecule and $q$ gives the number of excited phonons. The spin orientation $\sigma=\uparrow,\,\downarrow$ is only relevant for the singly-occupied molecule.  Since we consider a spin-degenerate orbital and spin-independent tunneling matrix elements $t_{L,R}(z)$, cf.~Eq.~\eqref{Hmol} and \eqref{Hmix}, there exists a symmetry between the two states $\ket{1\uparrow,q}$ and $\ket{1\downarrow,q}$. This allows for a notationally more simple treatment without reference to specific spin states by introducing appropriate spin factors into the transition rates, which account for the multiplicity of the $n=1$ level, cf.~Appendix \ref{app:spin}.

The operator $H_\text{mix}$ introduces transitions between the eigenstates $\ket{n,q}$, for which the rates are calculated via Fermi's golden rule in the next subsection. Subsequently, these are used to formulate the rate equations and the expression for the steady-state current.

\subsection{Transition rates}
We abbreviate the total rate for a transition $\ket{n,q}\rightarrow\ket{n',q'}$ by $W^{n\rightarrow n'}_{q\rightarrow q'}$. These total transition rates can be written as a product of a Fermi factor $f_a$ or $(1-f_a)$, which gives the probability for the availability of electrons or holes at the appropriate energy in lead $a$, and a ``bare transition rate" factor $\Gamma^{n\rightarrow n'}_{q\rightarrow q'}$  calculated by Fermi's golden rule. We denote the energy of the molecule in the state $\ket{n,q}$ by
\be
E^n_q= n(\varepsilon-eV_g) + U n(n-1)/2 + \hbar\omega(q+1/2).
\ee
\begin{figure}[ht]
	\begin{center}
		\includegraphics[width=0.7\columnwidth]{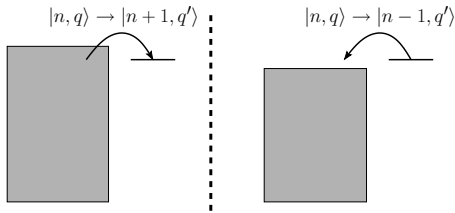}		
	\end{center}
	\caption{Sequential-tunneling processes, here schematically exemplified for tunneling between the molecule and the left lead.\label{fig:sequential}}
\end{figure}
Leading order perturbation theory yields sequential-tunneling processes $\ket{n,q}\rightarrow \ket{n\pm1,q'}$, cf.~Fig.~\ref{fig:sequential}, with the total rates
\begin{align}\label{sqrates1}
         W^{n\rightarrow (n+1)}_{ q\rightarrow q'} &= \sum_{a=L,R} f_a\left( E^{(n+1)}_{q'}-E^n_q\right) 						
         \Gamma^{n\rightarrow (n+1)}_{ q\rightarrow q';\,a}\;,\\\label{sqrates2}
         W^{n\rightarrow (n-1)}_{ q\rightarrow q'} &=\sum_{a=L,R} \left[1-f_a\left( 
         E^{n}_{q}-E^{(n-1)}_{q'}\right)\right] \Gamma^{n\rightarrow (n-1)}_{q\rightarrow q';\,a}.
      \end{align}
The bare transition rates $\Gamma$ are obtained by using Fermi's golden rule,
\begin{align}
\Gamma^{n\rightarrow (n-1)}_{q\rightarrow q';\,a}&= 
s^{n\rightarrow (n-1)}\frac{2\pi}{\hbar}\rho_a(E^n_q-E^{(n-1)}_{q'})\nonumber\\ &\qquad\times\abs{\boket{n-1,q',\epsilon^a}{H_\text{mix}}{n,q,0}}^2\\
&= s^{n\rightarrow (n-1)}\frac{2\pi}{\hbar}\rho_a(E^n_q-E^{(n-1)}_{q'})\abs{M^{n\rightarrow (n-1)}_{q\rightarrow q';\,a}}^2\;,\nonumber
\end{align}
and analogously
\be
\Gamma^{n\rightarrow (n+1)}_{q\rightarrow q';\,a}=s^{n\rightarrow (n+1)}\frac{2\pi}{\hbar}\rho_a(E^{(n+1)}_{q'}-E^{n}_{q})\abs{M^{n\rightarrow (n+1)}_{q\rightarrow q';\,a}}^2.
\ee
Here, $\rho_a$ denotes the density of states in lead $a$, and $s^{n\rightarrow m}$ denotes the spin factor, cf.~Appendix \ref{app:spin}.  In our calculations we assume $\rho_L=\rho_R=\text{const}$.
We note that due to our choice of symmetric tunneling matrix elements $t_L(z)=t_R(-z)$, cf.~Eq.~\eqref{telem}, all rates $\Gamma^{n\rightarrow (n\pm1)}_{q\rightarrow q'}$ are \textit{de facto} independent of the lead index $a$. The matrix elements $M^{n\rightarrow (n\pm1)}_{q\rightarrow q';\,a}$ can be expressed in terms of Laguerre polynomials and their detailed form depends on the phonon type, see Appendix \ref{sec:trates}.

\begin{figure}[ht]
	\begin{center}
	 \includegraphics[width=\columnwidth]{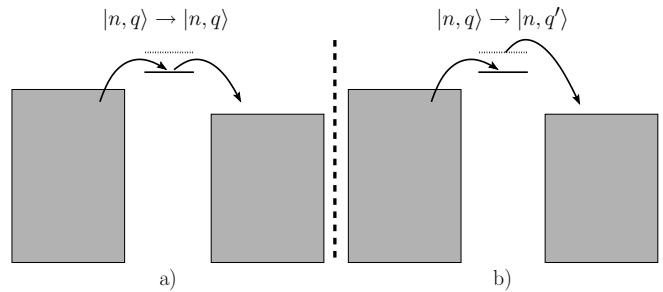}
		\end{center}
\caption{Cotunneling processes. a) Elastic cotunneling, b) inelastic cotunneling\label{fig:co}}
\end{figure}
Next-leading order perturbation theory generates cotunneling processes with a virtual intermediate molecule state.\cite{footnote}
 Cotunneling leaves the electron number on the molecule unchanged. By contrast, the phonon state can be changed, and in this case the process is called inelastic cotunneling. 
 If initial and final state of the molecule are identical, one speaks of elastic cotunneling, cf.~Fig.~\ref{fig:co}.
 In addition to the examples depicted in Fig.~\ref{fig:co}, where the electron is transferred from the left to the right lead, there are also contributions for the reverse process and for processes which involve tunneling back and forth between only one lead and the molecule. Thus, we abbreviate the total cotunneling rates by $W^{n\rightarrow n}_{q\rightarrow q';\,a\rightarrow b}$, where $a$ and $b$ denote either $L$ or $R$ for the left and right lead.

In order to obtain the total rates, one needs to sum over all lead energies $\epsilon^a$ weighted by Fermi functions for occupancy probabilities in the appropriate leads:
\begin{align}\label{corate}
&W^{n\rightarrow n}_{ q\rightarrow q';\, a\rightarrow b}\\
& \quad = \int d\epsilon^{a}\, \rho_a(\epsilon^{a}) f_a(\epsilon^{a})\left[1- f_{b}(\epsilon^{b})
\right]\Gamma^{n\rightarrow n}_{q\rightarrow q';a\rightarrow b}(\epsilon^{a}),\nonumber
\end{align}
where $\epsilon^{b}=\epsilon^a + E^n_q - E^n_{q'}$ as required by energy conservation. The bare cotunneling rates are given by

\begin{align}\label{gammaco1}
&\Gamma^{n\rightarrow n}_{q\rightarrow q';\,a\rightarrow b}(\epsilon^a)
 =s^{n\rightarrow n}\frac{2\pi}{\hbar} \rho_b(\epsilon^a - E^n_q + E^n_{q'})\\
& \times\abs{\sum_{q''} \frac{M^{n\rightarrow(n+1)}_{q\rightarrow q'';\,a}M^{(n+1)\rightarrow n}_{q''\rightarrow q';\,b}}{\epsilon^a+E^n_q-E^{n+1}_{q''}}+\frac{M^{n\rightarrow (n-1)}_{q\rightarrow q'';\,b}M^{(n-1)\rightarrow n}_{q''\rightarrow q';\,a}}{E^n_{q'}-\epsilon^{a}-E^{n-1}_{q''}}}^2.\nonumber
\end{align}
Here, the first term in the sum represents cotunneling processes with an electron virtually tunneling from a lead into the molecule and out again. Conversely, the second term contains those cotunneling processes in which an electron virtually tunnels out of the molecule and another electron tunnels back in subsequently. Again, the symmetry of the tunneling matrix elements, $t_L(z)=t_R(-z)$, leads to $\Gamma^{n\rightarrow n}_{q\rightarrow q';\,L\rightarrow R}=\Gamma^{n\rightarrow n}_{q\rightarrow q';\,R\rightarrow L}$ and $\Gamma^{n\rightarrow n}_{q\rightarrow q';\,L\rightarrow L}=\Gamma^{n\rightarrow n}_{q\rightarrow q';\,R\rightarrow R}$.

It has been pointed out  previously in the literature that due to the singularities of the bare transition rates \eqref{gammaco1} for cotunneling, the integrals \eqref{corate} for $W$ diverge at finite temperatures.\cite{averin,turek}
 Physically, this problem arises since we have assumed a well-defined energy for the intermediate virtual state. In reality, due to higher order tunneling effects, this intermediate state gains a finite width $\Gamma_i$, which leads to a regularization of the integrals. In the limit $\Gamma_i\rightarrow0$ one can derive a regularization scheme analogous to that presented by Turek and Matveev in Ref.~\onlinecite{turek}. This scheme consists of two steps: Firstly, introduce finite widths $\Gamma_i$ of the intermediate states by adding an imaginary part in the energy denominators. Secondly, in order to avoid double counting, subtract terms scaling as $1/\Gamma_i$, which correspond to sequential-tunneling contributions. In the $\Gamma_i\rightarrow0$ limit, the remaining integrals can be written as Cauchy principal-value integrals. Formally, this scheme can be extended to our case including phonons. However, this turns out to be a rather awkward procedure concerning the numerical treatment of the integrals. Instead, we find it useful to transform the integrals so that a power series expansion of the expressions with respect to $\Gamma_i$ is feasible. After the subtraction of all $1/\Gamma_i$ terms, the $\Gamma_i\rightarrow 0$ limit can be evaluated explicitly and we can express our results in terms of polygamma functions, cf.~Appendix \ref{app:reg}.

\subsection{Rate equations}
In the weak-coupling regime and for nondegenerate eigenstates of the molecule, it can be shown that the off-diagonal elements of the reduced density matrix of the molecule $\rho_\text{mol}$ are negligibly small.\cite{aleiner,blum} Therefore, the generalized master equations obtained via a density matrix approach reduce to simple rate equations. Writing $P^n_q(t)=\boket{n,q}{\rho_\text{mol}(t)}{n,q}$ for the probability that the molecule is in the state $\ket{n,q}$ at time $t$, one obtains:
\be
\partial P^n_q/\partial t = \sum_{n'\not=n}\sum_{q'\not=q} \left[ P^{n'}_{q'} W^{n'\rightarrow n}_{q'\rightarrow q} - P^{n}_{q} W^{n\rightarrow n'}_{q\rightarrow q'} \right].
\ee

We consider the leading and next-leading order contributions to the transition rates derived in the previous subsection.  Generally speaking, sequential tunneling is the dominant process close to the Coulomb peaks in $dI/dV$, i.e., whenever alignment of a molecular level with the Fermi energies of the leads permits tunneling:
\be k_BT\agt\min_{m=0,1} \abs{eV_g-\varepsilon-m U},\label{crit1}.\ee
Cotunneling plays the dominant role in the Coulomb valleys and for sufficiently low temperatures, i.e., when
\be k_BT \ll\min_{m=0,1} \abs{eV_g-\varepsilon-m U}.\label{crit2}\ee
Using the results for the transition rates, one obtains for the stationary case
			\be\begin{split}\label{master}
      0=\frac{\partial}{\partial t}P^n_q= &\sum_{q'} \left[ P^{(n-1)}_{q'}\; W^{(n-1)\rightarrow n}_{ q'\rightarrow q} 
                       + P^{(n+1)}_{q'}\;W^{(n+1)\rightarrow n}_{ q'\rightarrow q}\right.\\
                       &\qquad\qquad \left. -P^{n}_{q} \;W^{n\rightarrow (n+1)}_{ q\rightarrow q'}
                       - P^{n}_{q}\;W^{n\rightarrow (n-1)}_{ q\rightarrow q'}  \right]\\
             +&\sum_{q'\not=q}\left[ P^{n}_{q'}\;W^{n\rightarrow n}_{ q'\rightarrow q}- P^{n}_{q}\;W^{n\rightarrow n}_{       q\rightarrow q'} \right]\\
             -&\frac{1}{\tau}\left[{\textstyle P^n_q- P^\text{eq}_q \sum_{q'} P^n_{q'}   }\right],
      \end{split}\ee
where we have included an additional term, which takes into account relaxation of the phonons in the relaxation time approximation with relaxation time $\tau$. $P^\text{eq}_q=e^{-q\hbar\omega/k_BT}(1-e^{-\hbar\omega/k_BT})$ is the equilibrium probability distribution for the phonons alone. The various cotunneling contributions have been abbreviated by introducing
\be
W^{n\rightarrow n}_{ q\rightarrow q'}=\sum_{a=L,R}\sum_{b=L,R} W^{n\rightarrow n}_{q\rightarrow q';\,a\rightarrow b}.
\ee
The stationary-state rate equations in conjunction with the normalization condition $\sum_{n,q} P^n_q=1$ form an inhomogeneous system of linear equations whose solution gives the stationary probability distribution $P^n_q$ for a given voltage bias or temperature difference.

The stationary current is given by
\be\begin{split}\label{current}
I=&\sum_{n,q,q'}  P^{n}_{q}\left[ W^{n\rightarrow (n+1)}_{ q\rightarrow q';\,R}
				- W^{n\rightarrow (n-1)}_{ q\rightarrow q';\,R} \right]\\
+&\sum_{n,q,q'}P^{n}_{q} \left[ W^{n\rightarrow n}_{ q\rightarrow q';\, R\rightarrow L}
				-W^{n\rightarrow n}_{ q\rightarrow q';\, L\rightarrow R}\right].
\end{split}\ee
The first sum comprises all sequential-tunneling contributions, the second sum the cotunneling contributions.

Calculations of $IV$ characteristics based on this approach show phonon steps in the $IV$ curve for the sequential tunneling regime and phonon steps in $dI/dV$ for the cotunneling regime at low temperatures.\cite{aleiner, varma} The relaxation time approximation provides a means of analyzing situations between the two extremes of equilibrated and non-equilibrated phonons studied by Mitra \textit{et al}.\cite{aleiner}

\section{Thermopower\label{sec:thp}}
The thermopower, Eq.~\eqref{thpdef}, is calculated by considering the current through the molecule in the linear response regime, which is 
\be
I(V,\Delta T)=G V + G_T \Delta T +\mathcal{O}(V^2,\Delta T^2,V\Delta T),
\ee
where $G$ denotes the conductance and $G_T$ the thermal coefficient.
Hence, the thermopower can be written as
\be\label{thpower}
 S=\frac{G_T}{G}=\frac{G_T^\text{sq}+G_T^\text{co}}{G^\text{sq}+G^\text{co}},
 \ee
 where sequential tunneling and cotunneling contributions have been separated. We investigate sequential tunneling and cotunneling contributions to the thermopower and obtain expressions valid in the full crossover regime by means of the regularization scheme (Appendix \ref{app:reg}).
 
In order to obtain $G$ and $G_T$, we expand the current \eqref{current} in the bias voltage 
 $V=V_L-V_R$ and the temperature difference $\Delta T=T_L-T_R$ between the source and drain electrodes. 
Since $V$ and $\Delta T$ are in principle infinitesimal, we can conveniently choose the right electrode to have zero potential and temperature $T$. Accordingly, the left electrode has potential $V$ and temperature $T+\Delta T$. When expanding the current, one has to expand both the probabilities $P^n_q$ and the transition rates $W^{n\rightarrow n'}_{q\rightarrow q'}$.

We write the expansion for the transition rates and probabilities as
\be\label{expansion1}
W^{n\rightarrow n'}_{q\rightarrow q'}=w^{n\rightarrow n'}_{q\rightarrow q'} + \Delta T t^{n\rightarrow n'}_{q\rightarrow q'} + V v^{n\rightarrow n'}_{q\rightarrow q'}+\cdots\;
\ee
and
\be\label{expansion2}
P^n_q=\overline{P^n_q} + \Theta^n_q \Delta T + \Phi^n_q V+\cdots.
\ee
Here, $\overline{P^n_q}= 2^{\delta_{1,n}}\exp(-E^n_q/k_BT)/Z$ denotes the grand\-canonical probability distribution, and $Z=\sum_{n,q}2^{\delta_{1,n}}\exp(-E^n_q/k_BT)$ is the corresponding partition function. (The additional factor of $2^{\delta_{n,1}}$ takes the spin-degeneracy of the level $n=1$ into account.) The normalization condition for the deviations of $P^n_q$ from its equilibrium value is $0=\sum_{n,q}\Theta^n_q =\sum_{n,q}\Phi^n_q$.

By substituting the expansions \eqref{expansion1}, \eqref{expansion2} into the rate equations \eqref{master} and retaining only terms linear in $V$ and $\Delta T$, one obtains a new set of rate equations for the deviations $\Theta^n_q$ and $\Phi^n_q$, see Appendix \ref{app:newrate}, Eq.~\eqref{newrate}. These, in conjunction with the normalization conditions for $\Theta$ and $\Phi$, represent an inhomogeneous system of linear equations, whose solution yields the deviations $\Theta^n_q$ and $\Phi^n_q$.

Finally, the current \eqref{current} can be expanded in terms of $\Delta T$ and $V$:
\be\begin{split}\label{currexp}
I=&\sum_{n,q,q'}
      (\Delta T\Theta^n_q+V\Phi^n_q)\left[ 
      w^{n\rightarrow (n+1)}_{q\rightarrow q';\,R}-w^{n\rightarrow (n-1)}_{q\rightarrow q';\,R}
   \right]
  \\
  +&\sum_{n,q,q'}\bigg\{
   \overline{P^n_q}\left[
      \Delta T (t^{n\rightarrow n}_{q\rightarrow q';\,R\rightarrow L}-t^{n\rightarrow n}_{q\rightarrow q';\,L\rightarrow R})\right.\\
      &\left.\qquad\qquad+V (v^{n\rightarrow n}_{q\rightarrow q';\,R\rightarrow L}-v^{n\rightarrow n}_{q\rightarrow q';\,L \rightarrow R})
   \right]
  \bigg\}
\end{split}\ee
Note that the terms proportional to $\Theta^n_q$ and $\Phi^n_q$ in the expansion of the cotunneling contributions are absent since they come with the coefficient $\left[ 
      w^{n\rightarrow n}_{q\rightarrow q';\,R\rightarrow L}-w^{n\rightarrow n}_{q\rightarrow q';\,L \rightarrow R}
   \right]$, which vanishes due to the symmetry $\Gamma^{n\rightarrow n}_{q\rightarrow q';a\rightarrow b}(\epsilon)=\Gamma^{n\rightarrow n}_{q\rightarrow q';b\rightarrow a}(\epsilon)$.

(a) \emph{Sequential-tunneling contributions.}---In Eq.~\eqref{currexp} it was chosen to expand $I_R$ in $V$ and $\Delta T$. Due to the steady-state property $I=I_R=I_L$, an expansion in $I_L$ gives the same result and it turns out to be convenient to expand the expression $I=(I_L+I_R)/2$, which gives for the sequential-tunneling contributions
\begin{align}
I^\text{sq} =& \frac{1}{2} \sum_{n,q,q'}
      (\Delta T\Theta^n_q+V\Phi^n_q)\left[ 
      w^{n\rightarrow (n+1)}_{q\rightarrow q';\,R}-w^{n\rightarrow (n-1)}_{q\rightarrow q';\,R}\right.\nonumber\\
      &\qquad\qquad\qquad\left.+w^{n\rightarrow (n-1)}_{q\rightarrow q';\,L}-w^{n\rightarrow (n+1)}_{q\rightarrow q';\,L}
   \right]\nonumber\\
   +&\frac{1}{2} \sum_{n,q,q'}
      \overline{P^n_q}V\left[ 
      v^{n\rightarrow (n-1)}_{q\rightarrow q';\,L}-v^{n\rightarrow (n+1)}_{q\rightarrow q';\,L}\right]
      \\\nonumber
   +&\frac{1}{2} \sum_{n,q,q'}
      \overline{P^n_q}\Delta T\left[ 
      t^{n\rightarrow (n-1)}_{q\rightarrow q';\,L}-t^{n\rightarrow (n+1)}_{q\rightarrow q';\,L}\right].
\end{align}
Here, the first term remarkably vanishes due to the symmetry $w^{n\rightarrow (n\pm1)}_{q\rightarrow q';\,R} =w^{n\rightarrow (n\pm1)}_{q\rightarrow q';\,L}$. Therefore, one obtains the following sequential-tunneling contributions to the thermal coefficient $G_T$ and the conductance $G$,
\begin{align}\label{G_Tsq}
G_T^\text{sq} &= \frac{1}{2} \sum_{n,q,q'}
      \overline{P^n_q}\left[ 
      t^{n\rightarrow (n-1)}_{q\rightarrow q';\,L}-t^{n\rightarrow (n+1)}_{q\rightarrow q';\,L}\right],\\
      \label{Gsq}
G^\text{sq} &= \frac{1}{2} \sum_{n,q,q'}
      \overline{P^n_q}\left[ 
      v^{n\rightarrow (n-1)}_{q\rightarrow q';\,L}-v^{n\rightarrow (n+1)}_{q\rightarrow q';\,L}\right].
\end{align}
We point out that the so-obtained conductance and thermal coefficient do not depend on the probability deviations $\Theta^n_q$ and $\Phi^n_q$ any more. This is an important result since it allows for an analytic expression of the thermopower not involving an explicit solution of the rate equations, cf.~Appendix \ref{app:newrate}. (Expansions of $I_L$ and $I_R$ alone lead to expressions for $G^\text{sq}$ and $G^\text{sq}_T$, which do involve $\Theta^n_q$ and $\Phi^n_q$. We have also carried out calculations based on this approach by solving the rate equations for the probability deviations and find agreement with the results from Eqs.~\eqref{G_Tsq} and \eqref{Gsq}.)

(b) \emph{Cotunneling contributions}---The cotunneling contributions to thermal coefficient and conductance are 
\begin{align}\label{Gco_T}
G^\text{co}_T&=\sum_{n,q,q'}  \overline{P^n_q} (t^{n\rightarrow n}_{q\rightarrow q';\,R\rightarrow L}-t^{n\rightarrow n}_{q\rightarrow q';\,L\rightarrow R}),\\
\label{Gco}
G^\text{co}&=\sum_{n,q,q'}  \overline{P^n_q} (v^{n\rightarrow n}_{q\rightarrow q';\,R\rightarrow L}-v^{n\rightarrow n}_{q\rightarrow q';\,L\rightarrow R}).
\end{align}
In principle, any set of rate equations involving phononic excitations yields an infinite system of linear equations. In numerical calculations one makes use of the fact that transition matrix elements involving highly excited phonon states are typically very small. This allows for the introduction of a cutoff phonon number.

We find that the linear response quantities $G$ and $G_T$ do not depend on the relaxation time $\tau$. Mathematically, this corresponds to the result that the conductance and thermal coefficient do not involve the probability deviations $\Phi^n_q$ and $\Theta^n_q$, cmp.~Eqs.~\eqref{G_Tsq}--\eqref{Gco}. The physical reason for this is the following: In the $I\rightarrow 0$ limit the average time needed for one electron tunneling through the molecule becomes large compared to the relaxation time. Consequently, the initial state for any tunneling process corresponds to an equilibrium phonon state.

By substituting back Eqs.~\eqref{G_Tsq}--\eqref{Gco} into Eq.~\eqref{thpower}, we arrive at the following analytical expression for the thermopower:
\begin{widetext}
\be\label{theresult}
S=\frac{ \sum_{n,q,q'}
      \overline{P^n_q}\left[ 
      t^{n\rightarrow (n-1)}_{q\rightarrow q';\,L}-t^{n\rightarrow (n+1)}_{q\rightarrow q';\,L} +2t^{n\rightarrow n}_{q\rightarrow q';\,R\rightarrow L}-2t^{n\rightarrow n}_{q\rightarrow q';\,L\rightarrow R}\right]
      }{ \sum_{n,q,q'}
      \overline{P^n_q}\left[ 
      v^{n\rightarrow (n-1)}_{q\rightarrow q';\,L}-v^{n\rightarrow (n+1)}_{q\rightarrow q';\,L} +2v^{n\rightarrow n}_{q\rightarrow q';\,R\rightarrow L}-2v^{n\rightarrow n}_{q\rightarrow q';\,L\rightarrow R}\right]}.
\ee
\end{widetext}
This equation is our central result. It shows that even in the presence of phonons the thermopower can be expressed analytically through the equilibrium probability distribution $\overline{P^n_q}$  and the expansion coefficients of the transition rates evalutated at vanishing source drain voltage and temperature difference. In the following section the implications of Eq.~\eqref{theresult} will be discussed.

\section{Results\label{sec:res}}
\subsection{Sequential tunneling\label{sec:sqres}}
\begin{figure}
\vspace{0.2cm}
	\begin{center}
		\includegraphics[width=0.9\columnwidth]{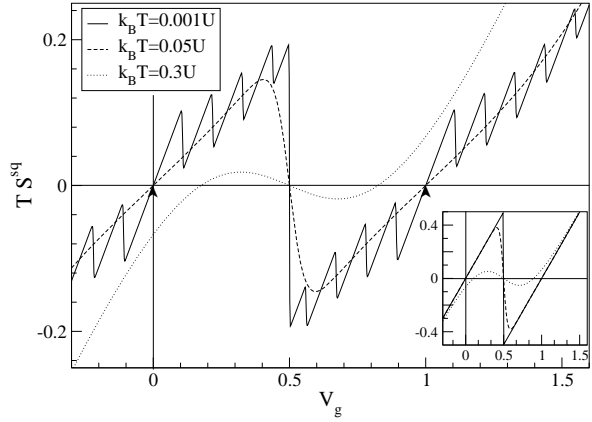}
	\end{center}
	\caption{Thermopower $S^\text{sq}$ times temperature as a function of gate voltage. The parameters are chosen as $\varepsilon=0$, $\hbar\omega=0.11U$. $V_g$ and $T\,S$ have units of $U/e$. Large: Vibrations with $\lambda=2$. Inset: Oscillations with $\xi_0=5$. (See text below Eq.~\protect\eqref{steps} for a discussion of the choice of parameters.) The positions of the corresponding Coulomb peaks in $\partial I/\partial V$ are marked with arrows. In contrast, the main features of the thermopower occur between the Coulomb peaks.\label{fig:Snew}}
\end{figure}

We first consider the results for pure sequential tunneling, postponing the discussion of the full thermopower due to both sequential and cotunneling to Sec.~\ref{sec:total}. We give numerical results for  the thermopower and present analytic expressions for the limiting case $U\rightarrow\infty$ and $T\rightarrow0$ below. Representative numerical results are shown in Fig.~\ref{fig:Snew}.

The dominant feature of the sequential-tunneling thermopower $S^\text{sq}$ is a large step at the gate voltage $V_g^*=(\varepsilon+U/2)/e$ for which the Fermi energy of the leads lies halfway in between the $\ket{1,0}$ and the $\ket{2,0}$ state. 
\begin{figure}[ht]
	\begin{center}
		\includegraphics[width=0.7\columnwidth]{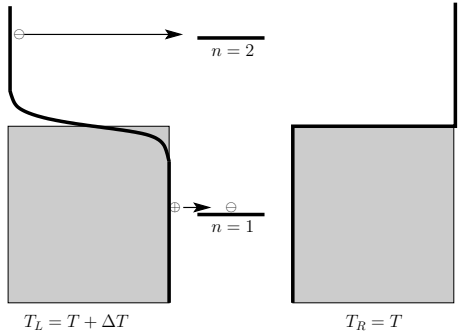}		
	\end{center}
	\caption{Electron-hole symmetry for the gate voltage $V_g^*=(\varepsilon+U/2)/e$.\label{fig:sym}}
\end{figure}
This situation is depicted in Fig.~\ref{fig:sym}. At $V_g=V_g^*$ the thermocurrent $G_T \Delta T$ vanishes due to electron-hole symmetry. An increased (decreased) gate voltage lowers (raises) the molecular levels with respect to the Fermi level, and therefore current is dominated by electrons (holes) flowing from the left to the right lead. Consequently, $G_T$ changes sign  at $V_g=V_g^*$. Moreover, away from the Coulomb peaks, sequential tunneling can only occur through the tails of the lead Fermi distributions due to energy conservation. Therefore, the sequential-tunneling  conductance and thermal coefficient fall off exponentially away from the Coulomb peak. At gate voltages close to $V_g^*$, their behavior can be estimated by
\begin{align}
G^\text{sq}&\sim \exp[-e(V_g^*-\varepsilon)/k_BT]\cosh[e(V_g^*-V_g)/k_BT],\\
G_T^\text{sq}&\sim \exp[-e(V_g^*-\varepsilon)/k_BT]\sinh[e(V_g^*-V_g)/k_BT].
\end{align}
Consequently, the sequential-tunneling thermopower $S^\text{sq}=G_T^\text{sq}/G^\text{sq}$ behaves as $\sim\tanh[e(V_g^*-V_g)/k_BT]$ in the vicinity of $V_g^*$, which develops a discontinuity in the limit $T\rightarrow 0$.

In addition to the electronic step, the results show smaller phonon steps with a distance of $\hbar\omega$ between adjacent steps. To understand slope, 
temperature dependence, and phonon step sizes of the  sequential-tunneling thermopower, we turn to the case $U\rightarrow\infty$. 

In the $U\rightarrow\infty$ limit, electronic double occupation of the molecule is forbidden, and the
 sequential-tunneling thermopower $S^\text{sq}=G^\text{sq}_T/G^\text{sq}$ reads
\begin{align}
&S^\text{sq}=\frac{\sum_{q,q'}P_q^\text{eq}(\overline{P^0}t^{0\rightarrow1}_{q\rightarrow q'}-\overline{P^1}t^{1\rightarrow0}_{q\rightarrow q'})}{\sum_{q,q'}P_q^\text{eq}(\overline{P^0}v^{0\rightarrow1}_{q\rightarrow q'}-\overline{P^1}v^{1\rightarrow0}_{q\rightarrow q'})}\nonumber
= \frac{V_g-\varepsilon/e}{T}\\
&-\frac{\sum_{q,q'}(\overline{P^0} P^\text{eq}_q + \overline{P^1} P^\text{eq}_{q'}) f_R'(E^1_{q'}-E^0_q)\hbar\omega(q'-q)\Gamma^{0\rightarrow1}_{q'\rightarrow q}}{Te\sum_{q,q'}(\overline{P^0} P^\text{eq}_q + \overline{P^1} P^\text{eq}_{q'}) f_R'(E^1_{q'}-E^0_q)\Gamma^{0\rightarrow1}_{q'\rightarrow q}}.
\end{align}
Thus, we find that the thermopower purely due to sequential tunneling roughly scales like $1/T$, which is in agreement with the quantum dot case.\cite{beenakker1} In the low temperature limit, the thermopower as a function of gate voltage develops a characteristic sawtooth behavior. The slope of the linear pieces is found to be $dS^\text{sq}/dV_g=1/T$. 

In the $T\rightarrow0$ limit, one obtains 
\begin{align}\label{Slimit}
\lim_{T\rightarrow0} T S^\text{sq}&=V_g-\varepsilon/e\\
&- \sgn(eV_g-\varepsilon)\frac{\sum_{q<\abs{eV_g-\varepsilon}/\hbar\omega}\hbar\omega q\, \Gamma^{0\rightarrow1}_{0\rightarrow q}}{e\sum_{q<\abs{eV_g-\varepsilon}/\hbar\omega} \Gamma^{0\rightarrow1}_{0\rightarrow q}},\nonumber
\end{align}
where the last term generates the step features by adding up higher phonon contributions for increasing gate voltages.
We can obtain the phononic step size $\Delta_Q$ of the $Q$th step of $T S^\text{sq}$ in the $T\rightarrow 0$ limit from Eq.~\eqref{Slimit},
\be\begin{split}
\Delta_Q=&\hbar\omega/e\,\Gamma^{0\rightarrow 1}_{0\rightarrow Q} \\
&\times\sum_{q=0}^Q(Q-q)\Gamma^{0\rightarrow 1}_{0\rightarrow q} \left[ \sum_{q=0}^Q\Gamma^{0\rightarrow 1}_{0\rightarrow q}\sum_{q=0}^{Q-1}\Gamma^{0\rightarrow 1}_{0\rightarrow q} \right]^{-1}.\label{steps}
\end{split}\ee
Here, $Q$ counts the discontinuities of $T S^\text{sq}$ starting at $\varepsilon-eV_G=0$ with increasing gate voltage.
$\Delta_Q$ depends on the step number $Q$ and on the coupling strength $\lambda$ or $\xi_0=z_0/\lambda_\text{osc}$ for vibrations or oscillations, respectively. Here, $\lambda_\text{osc}=(\hbar/M\omega_\text{osc})^{1/2}$ is the harmonic-oscillator length for oscillations and $M$ denotes the molecular mass.
\begin{figure}
	\begin{center}
		\includegraphics[width=0.8\columnwidth]{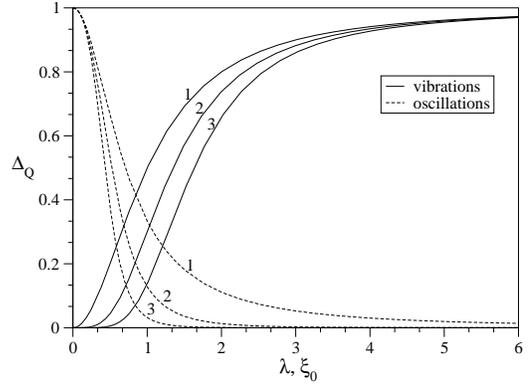}
	\end{center}
	\caption{Phonon step size $\Delta_Q$ in units of $\hbar\omega/e$ for the step numbers $Q=1,2,3$.	\label{fig:steps}}
\end{figure}

It is instructive to estimate typical values of the parameters $\xi_0$ and $\lambda$ for realistic systems. For $\xi_0$ we need to compare $z_0$ to the oscillator length $\lambda_\text{osc}$. An order of magnitude estimate yields $z_0\approx \hbar/(m_e W)^{1/2}$. Here, $W$ is the work function of the metal leads and is of the order of several eV. On the other hand, for a typical experiment\cite{park} oscillations occur on an energy scale of $1$--$10$ meV. This yields $\xi_0=z_0/\lambda_\text{osc}\gg1$. In this case, displacements of the molecule's CM are small on the scale of $z_0$, and therefore no significant shuttle effects can be expected.
 
Next we consider the vibrational coupling parameter $\lambda$. Let $r$ be the normal coordinate deviation from the equilibrium value $r_0$. An additional electron has to lowest order the effect of shifting the phonon potential curve by some distance $\Delta r$, so that the potential energy is now $\approx\frac{1}{2} M\omega_\text{vib}^2(r+n \Delta r)^2$. Hence, the electron-phonon coupling term is of the order of magnitude of $M\omega_\text{vib}^2r\Delta r= \Delta r/\lambda_\text{vib}\hbar\omega_\text{vib}(b+b^\dag)$ and therefore
$\lambda \approx \Delta r / \lambda_\text{vib}$.
Here, $\lambda_\text{vib}=(\hbar/M\omega_\text{vib})^{1/2}$ is the harmonic oscillator length corresponding to vibrations. There is not a general rule for how $\lambda_\text{vib}$ and $\Delta r$ compare so that $\lambda$ can in principle assume values both smaller and larger than $1$.

Due to the different behavior of the matrix elements for vibrations and oscillations, the phonon step size turns out to differ between those two cases as shown in Fig.~\ref{fig:steps}. For vibrational phonons the electron-phonon coupling becomes stronger for increasing $\lambda$. In the case of electron-phonon coupling for oscillations, the coupling gets stronger for decreasing $\xi_0=z_0/\lambda_\text{osc}$ (decreasing $z_0$ at fixed $\lambda_\text{osc}$ increases the position dependence of the hopping matrix elements $t(z)$). Thus, the plausible finding is that in both cases phonon step size increases with electron-phonon coupling strength. For oscillations the steps are rather small in the relevant regime of $\xi_0\gg1$. For vibrations they may be more pronounced.

\subsection{Results for the total thermopower\label{sec:total}}
The results for the thermopower discussed above arise from considering sequential-tunneling contributions only. However, if the Fermi levels are not aligned with a molecular level, sequential tunneling only occurs via electrons (or holes) in the tails of the Fermi distributions in the leads. In this case, the sequential-tunneling conductance $G^\text{sq}$ and thermal coefficient $G_T^\text{sq}$ are exponentially suppressed, and higher-order processes such as cotunneling may yield important contributions. 
Accordingly, sequential tunneling dominates in proximity to the aligned-levels configuration, and is suppressed most at the gate voltage $V_g^*$ at which the large electronic step in the sequential-tunneling thermopower occurs. In the latter range of gate voltages, cotunneling may give the dominant contributions to the thermopower. For this reason we have included the effect of cotunneling processes in the rate-equations approach, cf.~Sec.~\ref{sec:rate}. At temperatures $k_BT<\hbar\omega$ inelastic cotunneling can be neglected, and we find that the elastic cotunneling does not exhibit significant phonon structure.

\begin{figure}
\vspace*{0.47cm}
	\begin{center}
		\includegraphics[width=0.9\columnwidth]{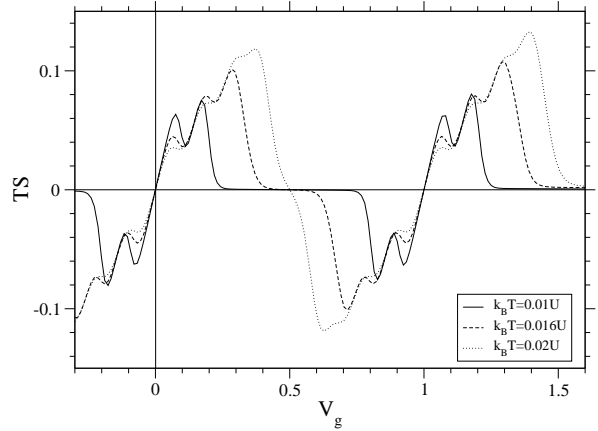}
			\end{center}
	\caption{Thermopower times temperature as a function of gate voltage for several temperatures. The parameter choices are: $\varepsilon=0$, $\hbar\omega=0.11U$, $\lambda=2$ (vibrations), $\alpha=10^{-9}$ (dimensionless coupling parameter defined in Eq.~\protect\eqref{alpha}).  $V_g$ and $TS$ are plotted in units of $U/e$.\label{fig:StotT}}
\end{figure}

Figure \ref{fig:StotT} exemplifies the behavior of the thermopower including both sequential and cotunneling as a function of gate voltage for several temperatures. 
Whether the total thermopower $S$ shows the sequential-tunneling phonon structure or whether it is mainly dominated by cotunneling contributions without significant phonon features, strongly depends on the choice of parameters. Firstly, step-like features can only be expected if the sequential-tunneling part develops pronounced steps. As discussed above, this depends on phonon type, phonon-coupling strength, and temperature. Only for temperatures well below $\hbar\omega/k_B$ one can expect any features, as can be seen from the smoothening of the phonon steps for increasing temperature in Fig.~\ref{fig:StotT}.

\begin{figure}
	\begin{center}
	\vspace*{0.3cm}
		\includegraphics[width=0.9\columnwidth]{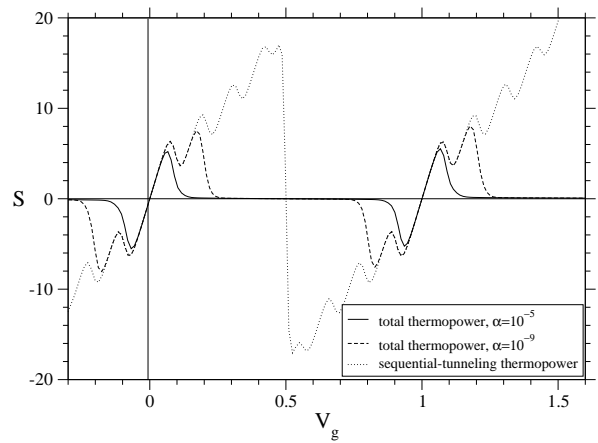}
			\end{center}
	\caption{Thermopower as a function of gate voltage at fixed temperature. The parameter choices are: $\varepsilon=0$, $\hbar\omega=0.11U$, $\lambda=2$ (vibrations), $T=0.01U/k_B$.  $V_g$ is plotted in units of $U/e$, $S$ in units of $k_B/e$.	\label{fig:Stot}}
\end{figure}

Secondly, temperature and the dimensionless coupling parameter,
\be\label{alpha}
 \alpha=\rho t_0^2/U,
\ee 
which arises in the rate equations and roughly describes the relative strength of cotunneling to sequential tunneling,\footnote{We measure the sequential and cotunneling rates  in natural units of $\rho t_0^2/\hbar$ and $\rho^2 t_0^4/(\hbar^2 U)$, respectively.} determine where the crossover between the sequential-tunneling and cotunneling regimes occurs. For illustration, Fig.~\ref{fig:Stot} shows the thermopower $S$ as a function of gate voltage at fixed temperature for two different coupling parameters $\alpha$ as well as the corresponding sequential-tunneling result for comparison.

The crossover between the sequential-tunneling and cotunneling regimes occurs in a rather small gate voltage range, cf.~Fig.~\ref{fig:StotT} and \ref{fig:Stot}, which allows one to identify  crossover gate voltages 
 $V_g^\text{xo}$.\footnote{Strictly speaking, the crossover ``point" is a region, and can be slightly different for $G$ and $G_T$.} Our results show that the crossover points roughly scale as $V_g^\text{xo}\sim T \ln \alpha^{-1}$, which is in agreement with corresponding results for quantum dots, cf., e.g., Ref.~\onlinecite{turek}.
 
This can be understood based on the following estimate of the crossover points $V_g^\text{xo}$. We assume that only a small gate-voltage region is dominated by sequential tunneling, $V_g^\text{xo}<V_g^*$. With $e\Delta V_g=\min_{m=0,1} \abs{eV_g-\varepsilon-mU}$ being the dominant activation energy for either electrons or holes, one can roughly estimate the sequential-tunneling conductance and  thermal coefficient by an activated behavior dependence,
\begin{equation}\label{Gapp1}
G^\text{sq},G^\text{sq}_T\sim  \exp (-e\Delta V_g/k_BT).
\end{equation}
While sequential-tunneling contributions therefore fall off exponentially with $\Delta V_g$, cotunneling contributions only show a  weak power-law dependence on the activation energy $e\Delta V_g$, and temperature $T$. To lowest order they may be approximated by a constant, 
\begin{equation}\label{Gapp2}
G^\text{co},G^\text{co}_T\sim \alpha b.
\end{equation}
Comparison of equations \eqref{Gapp1} and \eqref{Gapp2} yields as an estimate for the crossover gate voltage
\be\label{xo-est}
V_g^\text{xo}\approx k_BT (\ln \alpha^{-1} - \ln b)/e.
\ee
In this approximation, $V_g^\text{xo}$ increases linearly with temperature and decreases logarithmically with $\ln \alpha^{-1}$. The number of phonon steps (if present in the sequential-tunneling contribution) is given by $eV_g^\text{xo}/\hbar\omega$.

For the parameter choices of Fig.~\ref{fig:Stot}, we find that $b$ assumes values so that the  $\ln b$ term in Eq.~\eqref{xo-est} can be neglected. This leads to the following estimates of the crossover gate voltage: $V_g^\text{xo}\approx 0.1U/e$ for $\alpha=10^{-5}$, and $V_g^\text{xo}\approx 0.2U/e$ for $\alpha=10^{-9}$, which is in good agreement with the crossovers observed in Fig.~\ref{fig:Stot}.

We note that the thermopower attains rather small values in the cotunneling regime. This is plausible when reconsidering the effect of the electron-hole symmetry at $eV_g^*=(\varepsilon+U/2)$. Due to this symmetry, the thermopower must vanish at $V_g^*$. In the case of sequential tunneling, the exponentially suppressed current $G_T^\text{sq}\Delta T$ shows the breaking of this symmetry for small gate voltage deviations from $V_g^*$ rather abruptly (leading to the large steps in the sequential-tunneling thermopower $S^\text{sq}$ as a function of gate voltage). For cotunneling on the other hand, the thermal current is not exponentially suppressed, but roughly follows a power-like decrease with $\Delta V_g$. Therefore, breaking of the electron-hole symmetry is not as pronounced so that $S$ remains small in the cotunneling region centered around $V_g^*$.

\section{Summary\label{sec:sum}}
Using a model for electronic transport through a single spin-degenerate molecular orbital, and taking into account oscillational and vibrational phonons, we have calculated the thermopower of a single-molecule device in the regime of weak molecule-lead coupling. In contrast to $IV$ measurements, the thermopower provides a means of extracting information about electronic and phononic excitations and the nature of the electron-phonon coupling in a linear response measurement. Therefore, it may have advantages over the more conventional $IV$ characteristic, which necessarily involves nonequilibrium effects, and which, at large voltages, may even affect symmetry and structure of the molecule itself.

We have  found that sequential-tunneling contributions yield a characteristic sawtooth behavior of the thermopower as a function of gate voltage for low temperatures, which shows structure due to electronic and phononic excitations. It has been shown that due to the different nature of electron-phonon coupling for oscillations and vibrations, characteristic differences in the phonon step size arise in the sequential-tunneling thermopower. Analytical expressions for the phonon step size have been derived in the limit of strong Coulomb blockade, which show that, for realistic parameters, phonon steps can be expected to be more pronounced for vibrations than for oscillations.

Away from the Coulomb peaks, cotunneling dominates over sequential tunneling.\cite{footnote}
By considering cotunneling contributions obtained by means of a regularization scheme, we have found that the (elastic) cotunneling regime does not show significant structure due to phononic excitations. We have investigated the crossover regime and have given an estimate for the gate voltage at which the crossover takes place. It has been shown that the phononic structure exhibited by the sequential-tunneling contributions is retained in the thermopower if $k_BT\ll\hbar\omega$ and $k_BT\ln \alpha^{-1} > \hbar\omega$, i.e., (1) the temperature is low enough so that the phononic structure is not blurred out, and (2) the dimensionless coupling parameter $\alpha$ is so small that the crossover gate voltage is high enough to allow for at least one phonon feature in the thermopower.

\begin{acknowledgments}
We thank I. Bar-Joseph for suggesting an investigation of thermoelectric effects in single-molecule devices and C. Timm for useful discussions. Two of us (FvO and JK) thank the Weizmann Institute for hospitality (supported through the LSF and the Einstein Center). This work was supported in part by SFB 290, the Junge Akademie (FvO), as well as the Minerva Foundation, and DIP under Grant No. C-7.1 (YO).
\end{acknowledgments}

\appendix
\section{Spin factors\label{app:spin}}
In our model we consider transport through one spin-degenerate molecular orbital. Including the phononic excitations, a basis of the corresponding Hilbert space is given by the states 
\be
\ket{0,q},\qquad \ket{1\uparrow,q},\qquad \ket{1\downarrow,q},\qquad \ket{2,q}.
\ee
Due to the spin-degeneracy and the spin-independent tunneling matrix elements, one can make use of the resulting symmetry and obtain a description without explicit reference to spin indices. To see this, we write down the rate equations including spin indices,
\begin{align}\label{spinrate}
&\frac{\partial}{\partial t}P^{n\sigma}_q\nonumber\\
 &=\sum_{q',\sigma'} \left[ P^{(n-1)\sigma'}_{q'}\; \tilde{W}^{(n-1)\sigma'\rightarrow n\sigma}_{ q'\rightarrow q} 
                       + P^{(n+1)\sigma'}_{q'}\;\tilde{W}^{(n+1)\sigma'\rightarrow n\sigma}_{ q'\rightarrow q}\right.\nonumber\\
                       &\qquad\qquad \left. -P^{n\sigma}_{q} \;\tilde{W}^{n\sigma\rightarrow (n+1)\sigma'}_{ q\rightarrow q'}
                       - P^{n\sigma}_{q}\;\tilde{W}^{n\sigma\rightarrow (n-1)\sigma'}_{ q\rightarrow q'}  \right]\nonumber\\
             &\quad+\sum_{q'\not=q}\left[ P^{n\sigma}_{q'}\;\tilde{W}^{n\sigma\rightarrow n\sigma}_{ q'\rightarrow q}- P^{n\sigma}_{q}\;\tilde{W}^{n\sigma\rightarrow n\sigma}_{q\rightarrow q'}\right]\\
             &\quad+\sum_{q', \sigma'\not=\sigma}\left[ P^{n\sigma'}_{q'}\;\tilde{W}^{n\sigma'\rightarrow n\sigma}_{ q'\rightarrow q}- P^{n\sigma}_{q}\;\tilde{W}^{n\sigma\rightarrow n\sigma'}_{q\rightarrow q'}\right],\nonumber
\end{align}
where it is implied that $\sigma=0$ for $n=0,2$ and $\sigma=\uparrow,\downarrow$ for $n=1$. Here, $\tilde{W}$ denotes total transition rates between specific spin states.

By defining $P^{1}_q\equiv P^{1\uparrow}_q+P^{1\downarrow}_q$ and $W^{(1\pm1)\rightarrow 1}_{q\rightarrow q'}\equiv W^{(1\pm1)\rightarrow 1\uparrow }_{q\rightarrow q'}+W^{(1\pm1)\rightarrow 1\downarrow}_{q\rightarrow q'}$, the sequential-tunneling contributions to the rate equations can be cast into the form of Eq.~\eqref{master}. Accordingly, the sequential-tunneling spin factors are given by
\be s^{1\rightarrow 0}=s^{1\rightarrow 2}=1, \qquad s^{0\rightarrow 1}=s^{2\rightarrow 1}=2. \ee 

For cotunneling transitions $\tilde{W}^{00\rightarrow 00}_{q\rightarrow q'}$ and $\tilde{W}^{20\rightarrow 20}_{q\rightarrow q'}$, one has to take into account that the virtual intermediate state is now spin-degenerate, which leads to a factor of $4$ due to the coherent sum, cf.~Eq.~\eqref{gammaco1}. For cotunneling transitions with $n=1$, the intermediate state has no spin-degeneracy, but the rate equations contain an additional spin-flip channel, cf.~Eq.~\eqref{spinrate}, leading to a factor of $2$. Therefore, by defining the cotunneling spin factors by
\be s^{1\rightarrow 1}=2, \qquad s^{0\rightarrow 0}=s^{2\rightarrow 2}=4, \ee
the rate equations involving spin, Eq.~\eqref{spinrate}, can be transformed into the set of rate equations in Eq.~\eqref{master}.

\section{Matrix elements\label{sec:trates}}

(a) \emph{Matrix elements for oscillations}.---The matrix elements $M^{n\rightarrow m}_{q\rightarrow q';\,a}$ for oscillations are given by: 
{\allowdisplaybreaks
\begin{align}
&M^{n\rightarrow (n\pm 1)}_{q\rightarrow q';\,L}=t_0 \boket{q'}{\exp[- z/z_0]}{q}\\
&\quad\textstyle= t_0 \left( \frac{2^{q_2-q_1}\,q_1!}{q_2!} \right)^{1/2} \left( - \frac{1}{2\xi_0} \right)^{q_2-q_1} e^\frac{1}{4\xi_0^2} \, L_{q_1}^{q_2-q_1}(\textstyle-\frac{1}{2\xi_0^2})\nonumber\\
&M^{n\rightarrow (n\pm 1)}_{q\rightarrow q';\,R}=t_0 \boket{q'}{\exp[+ z/z_0]}{q}\\
&\quad\textstyle= t_0 \left( \frac{2^{q_2-q_1}\,q_1!}{q_2!} \right)^{1/2} \left( + \frac{1}{2\xi_0} \right)^{q_2-q_1} e^\frac{1}{4\xi_0^2} \, L_{q_1}^{q_2-q_1}(\textstyle-\frac{1}{2\xi_0^2})\nonumber
\end{align}}
where $q_1=\min\{q,q'\}$, $q_2=\max\{q,q'\}$, and $\xi_0=z_0/\lambda_\text{osc}$. Finally, $L^n_m(x)$ denotes the generalized Laguerre polynomial. Note that there is no dependence on whether the final electronic state is $n+1$ or $n-1$.

(b) \emph{Matrix elements for vibrations}.---For vibrations one obtains, cf.~Ref.~\onlinecite{aleiner},
\be\begin{split}
&M^{n\rightarrow (n-1)}_{q\rightarrow q';\,a}= t_0\big\langle\, q' \,\big|\, e^{-\lambda(b^\dag-b)}\,\big|\, q\,\big\rangle=t_0\left( q_1!/q_2! \right)^{1/2}\\
&\quad \times  \lambda^{q_2-q_1}e^{-\lambda^2/2} \, L_{q_1}^{q_2-q_1}\left(\lambda^2 \right)\times
\begin{cases}
(-1)^{q'-q} & \text{for } q'\ge q\\
1 & \text{for } q'< q,
\end{cases}
 \end{split}\ee 
where again $q_1=\min\{q,q'\}$, $q_2=\max\{q,q'\}$. Note that there is no dependence on the lead index $a$. The corresponding matrix element for a transition $n\rightarrow n+1$ can be obtained by using
\be
M^{n\rightarrow (n-1)}_{q\rightarrow q';\,a}=M^{n\rightarrow (n+1)}_{q'\rightarrow q;\,a}\,
\ee

See Secs.~\ref{sec:model} and \ref{sec:sqres} for definitions and a discussion of the parameters $\lambda_\text{osc}$, $z_0$ and $\lambda$.

\begin{widetext}
\section{Regularization scheme\label{app:reg}}
We sketch the regularization scheme that we apply in order to extract the cotunneling contributions from the diverging next-leading order perturbation theory. The integrals $w$, $t$, and $v$ for cotunneling, cf.~Eq.~\eqref{expansion1}, contain factors of Fermi functions and their derivatives, energy factors and the bare transition rates. Our starting point is to introduce a finite width for the intermediate state. The bare transition rates then assume the following form:
\[\begin{split}
\Gamma^{n\rightarrow n}_{q\rightarrow q';\,a\rightarrow b}&\sim\abs{\sum_{k=0}^\infty\left( \textstyle\frac{A_k}{\epsilon-E_k +\ii\Gamma_k} +\frac{B_k}{\epsilon-E'_k +\ii\Gamma_k} \right)}^2\\
 &=
\sum_k \left[ \abs{\textstyle\frac{A_k}{\epsilon-E_k +\ii\Gamma_k}}^2+\abs{\textstyle\frac{B_k}{\epsilon-E'_k +\ii\Gamma_k}}^2\right]
+2 \rre \sum_{q}\sum_{k<q}\left[ \textstyle\frac{A_k}{\epsilon-E_k +\ii\Gamma_k}\frac{A_q}{\epsilon-E_q -\ii\Gamma_q}
+\frac{B_k}{\epsilon-E'_k +\ii\Gamma_k}\frac{B_q}{\epsilon-E'_q -\ii\Gamma_q}\right]\\
&\qquad+2\rre \sum_q \sum_k \textstyle\frac{A_k}{\epsilon-E_k +\ii\Gamma_k}\frac{B_q}{\epsilon-E'_q -\ii\Gamma_q}
\end{split}\]
Accordingly, we need to consider several types of integrals. In the following, we always evaluate the integrals in the $\Gamma\rightarrow0$ limit. The notation $``-\mathcal{O}(1/\Gamma)"$ indicates that terms proportional to $1/\Gamma$ have been subtracted before carrying out the limit. These terms correspond to sequential-tunneling contributions, cf.~Ref.~\onlinecite{varma}.
\begin{align}
\begin{split}
I&=\int d\epsilon\, f(\epsilon-E_1)f(\epsilon-E_2)\textstyle\frac{1}{\epsilon-\epsilon_1-\ii\Gamma_1}\frac{1}{\epsilon-\epsilon_2+\ii\Gamma_2}\\
&=\frac{1}{\epsilon_1-\epsilon_2}\bigg\{\ii \pi+ n_B(E_1-E_2) \left[ -\psi(1/2+\ii\beta[E_1-\epsilon_1]/2\pi) +\psi(1/2-\ii\beta[E_1-\epsilon_2]/2\pi) \right]\\
&\qquad\qquad\qquad+ n_B(E_2-E_1) \left[ -\psi(1/2+\ii\beta[E_2-\epsilon_1]/2\pi) +\psi(1/2-\ii\beta[E_2-\epsilon_2]/2\pi) \right]\bigg\}
\end{split}\\
\begin{split}
J&=\int d\epsilon\, f(\epsilon-E_1)f(\epsilon-E_2)\frac{1}{(\epsilon-E)^2+\Gamma^2}+``-\mathcal{O}(1/\Gamma)"\\
&=\frac{\beta}{2\pi}\left\{ n_B(E_1-E_2) \iim \psi^{(1)}\left(1/2+\ii\beta[E-E_1]/2\pi\right) +n_B(E_2-E_1) \iim \psi^{(1)}\left(1/2+\ii\beta[E-E_2]/2\pi\right)\right\}
\end{split}
\end{align}
Here, $\psi^{(n)}(x)$ denotes the polygamma function of order $n$, cf.~Ref.~\onlinecite{abramo}, and $n_B(x)=(\exp(x/k_BT)-1)^{-1}$ is the Bose function.
By evaluating $I$ ($J$) in the limit $E_2\rightarrow\infty$, one obtains an expression for the integral $I'$ ($J'$)  with only one Fermi factor. 
{\allowdisplaybreaks
\begin{align}
\begin{split}
K&=\int d\epsilon\, f'(\epsilon-E_1)f(\epsilon-E_2)\textstyle\frac{1}{\epsilon-\epsilon_1-\ii\Gamma_1}\frac{1}{\epsilon-\epsilon_2+\ii\Gamma_2}=-\partial I/\partial E_1
\end{split}\\
\begin{split}
L&=\int d\epsilon\, f'(\epsilon-E_1)f(\epsilon-E_2)\frac{1}{(\epsilon-E)^2+\Gamma^2}
=-\partial J/\partial E_1
\end{split}\\
M&=\int d\epsilon\,(\epsilon-E_1) f'(\epsilon-E_1)f(\epsilon-E_2)\textstyle\frac{1}{\epsilon-\epsilon_1-\ii\Gamma_1}\frac{1}{\epsilon-\epsilon_2+\ii\Gamma_2}=G+(\epsilon_2-E_1)K\\
M'&=\int d\epsilon\,
(\epsilon-E_1)f'(\epsilon-E_1)\left[1-f(\epsilon-E_2)\right]\textstyle\frac{1}{\epsilon-\epsilon_1-\ii\Gamma_1}\frac{1}{\epsilon-\epsilon_2+\ii\Gamma_2}=\beta\,\partial/\partial \beta\, I'-M\\
N&=\int d\epsilon\, (\epsilon-E_1)f'(\epsilon-E_1)f(\epsilon-E_2)\frac{1}{(\epsilon-E)^2+\Gamma^2}+``-\mathcal{O}(1/\Gamma)"=\rre G +(E-E_1)L\\
N'&=\int d\epsilon\, (\epsilon-E_1)f'(\epsilon-E_1)\left[1-f(\epsilon-E_2)\right]\frac{1}{(\epsilon-E)^2+\Gamma^2}+``-\mathcal{O}(1/\Gamma)"=\beta\,\partial/\partial \beta\, J'-N\\
G&=\int d\epsilon\,  f'(\epsilon-E_1)f(\epsilon-E_2)\frac{1}{\epsilon-\epsilon_1-\ii\Gamma}\\
&= \frac{\ii\beta}{2\pi} n_B(E_1-E_2) \psi^{(1)}(1/2+\ii\beta(E_1-\epsilon_1)/2\pi)
 +n_B'(E_1-E_2)\left[ \psi(1/2+\ii\beta (E_1-\epsilon_1)/2\pi)-\psi(1/2+\ii\beta (E_2-\epsilon_1)/2\pi)\right]\nonumber
\end{align}}

\section{Rate equations for $\Theta^n_q$, $\Phi^n_q$; Expansion coefficients\label{app:newrate}}
The rate equations for the deviations $\Theta^n_q$ and $\Phi^n_q$ from the equilibrium probability have the following form:
\be\begin{aligned}\label{newrate}
0=&\quad\sum_{q'}\bigg[\overline{P^{n-1}_{q'}}(\Delta T t^{(n-1)\rightarrow n}_{q'\rightarrow q}+V v^{(n-1)\rightarrow n}_{q'\rightarrow q})
+\overline{P^{n+1}_{q'}}(\Delta T t^{(n+1)\rightarrow n}_{q'\rightarrow q}+V v^{(n+1)\rightarrow n}_{q'\rightarrow q})
  -\overline{P^{n}_{q}}(\Delta T t^{n\rightarrow (n+1)}_{q\rightarrow q'}+V v^{n\rightarrow (n+1)}_{q\rightarrow q'}\\
&\qquad +\Delta T t^{n\rightarrow (n-1)}_{q\rightarrow q'}+V v^{n\rightarrow (n-1)}_{q\rightarrow q'})+(\Delta T \Theta^{n-1}_{q'}+V \Phi^{n-1}_{q'})w^{(n-1)\rightarrow n}_{q'\rightarrow q}
+(\Delta T \Theta^{n+1}_{q'}+V \Phi^{n+1}_{q'})w^{(n+1)\rightarrow n}_{q'\rightarrow q}\\
&\qquad -(\Delta T \Theta^{n}_{q}+V \Phi^{n}_{q})(w^{n\rightarrow (n+1)}_{q\rightarrow q'}+w^{n\rightarrow (n-1)}_{q\rightarrow q'})\bigg]\\
&+\sum_{q'\not=q}\bigg[ \overline{P^{n}_{q'}}(\Delta T t^{n\rightarrow n}_{q'\rightarrow q}+V v^{n\rightarrow n}_{q'\rightarrow q})-\overline{P^{n}_{q}}(\Delta T t^{n\rightarrow n}_{q\rightarrow q'}+V v^{n\rightarrow n}_{q\rightarrow q'})
+(\Delta T \Theta^{n}_{q'}+V \Phi^{n}_{q'})w^{n\rightarrow n}_{q'\rightarrow q}
-(\Delta T \Theta^{n}_{q}+V \Phi^{n}_{q})w^{n\rightarrow n}_{q\rightarrow q'}\bigg]\\
&-\frac{1}{\tau}\bigg[ \Delta T \Theta^n_q+V\Phi^n_q - P^\text{eq}_q\sum_{q'} (\Delta T \Theta^n_{q'} + V \Phi^n_{q'}) \bigg]
\end{aligned}\ee
By expanding \eqref{sqrates1} and \eqref{sqrates2} one obtains the following expressions for the expansion coefficients of the sequential-tunneling rates $W^{n\rightarrow(n\pm1)}_{q\rightarrow q'}=\sum_a W^{n\rightarrow(n\pm1)}_{q\rightarrow q';\,a}$:
\begin{align}
w^{n\rightarrow(n+1)}_{q\rightarrow q';\,L}&=w^{n\rightarrow(n+1)}_{q\rightarrow q';\,R}= f_R(E^{n+1}_{q'}-E^n_q) \,\Gamma^{n\rightarrow(n+1)}_{q\rightarrow q'},\\
w^{n\rightarrow(n-1)}_{q\rightarrow q';\,L}&=w^{n\rightarrow(n-1)}_{q\rightarrow q';\,R}= \left[ 1-f_R(E^{n}_{q}-E^{n-1}_{q'})\right] \Gamma^{n\rightarrow(n-1)}_{q\rightarrow q'},\\
t^{n\rightarrow(n+1)}_{q\rightarrow q';\,L}&=-(E^{n+1}_{q'}-E^n_q)/T\,f_R'(E^{n+1}_{q'}-E^n_q)\, \Gamma^{n\rightarrow(n+1)}_{q\rightarrow q'},&t^{n\rightarrow(n+1)}_{q\rightarrow q';\,R}&=0,\\
t^{n\rightarrow(n-1)}_{q\rightarrow q';\,L}&=+(E^n_q-E^{n-1}_{q'})/T\,f_R'(E^n_q-E^{n-1}_{q'})\, \Gamma^{n\rightarrow(n-1)}_{q\rightarrow q'}, &t^{n\rightarrow(n-1)}_{q\rightarrow q';\,R}&=0,\\
v^{n\rightarrow(n+1)}_{q\rightarrow q';\,L}&=+e\,f_R'(E^{n+1}_{q'}-E^n_q)\, \Gamma^{n\rightarrow(n+1)}_{q\rightarrow q'}, &v^{n\rightarrow(n+1)}_{q\rightarrow q';\,R}&=0,\\
v^{n\rightarrow(n-1)}_{q\rightarrow q';\,L}&=-e\,f_R'(E^n_q-E^{n-1}_{q'})\, \Gamma^{n\rightarrow(n-1)}_{q\rightarrow q'},&v^{n\rightarrow(n-1)}_{q\rightarrow q';\,R}&=0.
\end{align}
Similarly, the expansion of the cotunneling rates \eqref{corate} yields
\allowdisplaybreaks
\begin{align}
w^{n\rightarrow n}_{q\rightarrow q';\,a\rightarrow b}&=\;\rho \int d\epsilon\, f_R(\epsilon)\left[1-f_R(\epsilon+E^n_q-E^n_{q'})\right]\Gamma^{n\rightarrow n}_{q\rightarrow q';\,a\rightarrow b}(\epsilon)\\
t^{n\rightarrow n}_{q\rightarrow q';\,L\rightarrow R}&=-\rho\int d\epsilon\, \frac{\epsilon}{T} f_R'(\epsilon)\left[1-f_R(\epsilon+E^n_q-E^n_{q'})\right] \Gamma^{n\rightarrow n}_{q\rightarrow q';\,L\rightarrow R}(\epsilon)\\
t^{n\rightarrow n}_{q\rightarrow q';\,R\rightarrow L}&=\;\rho\int d\epsilon\, \frac{\epsilon+E^n_q-E^n_{q'}}{T} f_R(\epsilon)f_R'(\epsilon+E^n_q-E^n_{q'})\, \Gamma^{n\rightarrow n}_{q\rightarrow q';\,R\rightarrow L}(\epsilon)\\
t^{n\rightarrow n}_{q\rightarrow q';\,L\rightarrow L}&=\;\rho\int d\epsilon\,\left( {\textstyle\frac{\epsilon+E^n_q-E^n_{q'}}{T}} f_R(\epsilon)f_R'(\epsilon+E^n_q-E^n_{q'})-{\textstyle\frac{\epsilon}{T}} f_R'(\epsilon)\left[1-f_R(\epsilon+E^n_q-E^n_{q'})\right]\right)\Gamma^{n\rightarrow n}_{q\rightarrow q';\,L\rightarrow L}(\epsilon)\\
t^{n\rightarrow n}_{q\rightarrow q';\,R\rightarrow R}&=\;0, \qquad\qquad v^{n\rightarrow n}_{q\rightarrow q';\,R\rightarrow R}=\;0\\
v^{n\rightarrow n}_{q\rightarrow q';\,L\rightarrow R}&=-e\rho\int d\epsilon\, f_R'(\epsilon)\left[1-f_R(\epsilon+E^n_q-E^n_{q'})\right] \Gamma^{n\rightarrow n}_{q\rightarrow q';\,L\rightarrow R}(\epsilon)\\
v^{n\rightarrow n}_{q\rightarrow q';\,R\rightarrow L}&=\;e\rho\int d\epsilon\, f_R(\epsilon)f_R'(\epsilon+E^n_q-E^n_{q'})\, \Gamma^{n\rightarrow n}_{q\rightarrow q';\,R\rightarrow L}(\epsilon)\\
v^{n\rightarrow n}_{q\rightarrow q';\,L\rightarrow L}&=\;e\rho\int d\epsilon\,\left(f_R(\epsilon)f_R'(\epsilon+E^n_q-E^n_{q'})- f_R'(\epsilon)\left[1-f_R(\epsilon+E^n_q-E^n_{q'})\right]\right)\Gamma^{n\rightarrow n}_{q\rightarrow q';\,L\rightarrow L}(\epsilon)
\end{align}

\end{widetext}


\begin{thebibliography}{30}
\expandafter\ifx\csname natexlab\endcsname\relax\def\natexlab#1{#1}\fi
\expandafter\ifx\csname bibnamefont\endcsname\relax
  \def\bibnamefont#1{#1}\fi
\expandafter\ifx\csname bibfnamefont\endcsname\relax
  \def\bibfnamefont#1{#1}\fi
\expandafter\ifx\csname citenamefont\endcsname\relax
  \def\citenamefont#1{#1}\fi
\expandafter\ifx\csname url\endcsname\relax
  \def\url#1{\texttt{#1}}\fi
\expandafter\ifx\csname urlprefix\endcsname\relax\def\urlprefix{URL }\fi
\providecommand{\bibinfo}[2]{#2}
\providecommand{\eprint}[2][]{\url{#2}}

\bibitem[{\citenamefont{Smit et~al.}(2002)\citenamefont{Smit, Noat, Untiedt,
  Lang, van Hemert, and van Ruitenbeek}}]{ruitenbeek}
\bibinfo{author}{\bibfnamefont{R.~H.~M.} \bibnamefont{Smit}},
  \bibinfo{author}{\bibfnamefont{Y.}~\bibnamefont{Noat}},
  \bibinfo{author}{\bibfnamefont{C.}~\bibnamefont{Untiedt}},
  \bibinfo{author}{\bibfnamefont{N.~D.} \bibnamefont{Lang}},
  \bibinfo{author}{\bibfnamefont{M.~C.} \bibnamefont{van Hemert}},
  \bibnamefont{and} \bibinfo{author}{\bibfnamefont{J.~M.} \bibnamefont{van
  Ruitenbeek}}, \bibinfo{journal}{Nature} \textbf{\bibinfo{volume}{419}},
  \bibinfo{pages}{906} (\bibinfo{year}{2002}).

\bibitem[{\citenamefont{Porath et~al.}(2000)\citenamefont{Porath, Bezryadin,
  de~Vries, and Dekker}}]{porath}
\bibinfo{author}{\bibfnamefont{D.}~\bibnamefont{Porath}},
  \bibinfo{author}{\bibfnamefont{A.}~\bibnamefont{Bezryadin}},
  \bibinfo{author}{\bibfnamefont{S.}~\bibnamefont{de~Vries}}, \bibnamefont{and}
  \bibinfo{author}{\bibfnamefont{C.}~\bibnamefont{Dekker}},
  \bibinfo{journal}{Nature} \textbf{\bibinfo{volume}{403}},
  \bibinfo{pages}{635} (\bibinfo{year}{2000}).

\bibitem[{\citenamefont{Reichert et~al.}(2002)\citenamefont{Reichert, Ochs,
  Beckmann, Weber, Mayor, and v.~L{\"o}hneysen}}]{weber}
\bibinfo{author}{\bibfnamefont{J.}~\bibnamefont{Reichert}},
  \bibinfo{author}{\bibfnamefont{R.}~\bibnamefont{Ochs}},
  \bibinfo{author}{\bibfnamefont{D.}~\bibnamefont{Beckmann}},
  \bibinfo{author}{\bibfnamefont{H.~B.} \bibnamefont{Weber}},
  \bibinfo{author}{\bibfnamefont{M.}~\bibnamefont{Mayor}}, \bibnamefont{and}
  \bibinfo{author}{\bibfnamefont{H.}~\bibnamefont{v.~L{\"o}hneysen}},
  \bibinfo{journal}{Phys. Rev. Lett.} \textbf{\bibinfo{volume}{88}},
  \bibinfo{pages}{176804} (\bibinfo{year}{2002}).

\bibitem[{\citenamefont{Chen et~al.}(1999)\citenamefont{Chen, Reed, Rawlett,
  and Tour}}]{chen}
\bibinfo{author}{\bibfnamefont{J.}~\bibnamefont{Chen}},
  \bibinfo{author}{\bibfnamefont{M.~A.} \bibnamefont{Reed}},
  \bibinfo{author}{\bibfnamefont{A.~M.} \bibnamefont{Rawlett}},
  \bibnamefont{and} \bibinfo{author}{\bibfnamefont{J.~M.} \bibnamefont{Tour}},
  \bibinfo{journal}{Science} \textbf{\bibinfo{volume}{286}},
  \bibinfo{pages}{1550} (\bibinfo{year}{1999}).

\bibitem[{\citenamefont{Dekker}(1999)}]{dekker}
\bibinfo{author}{\bibfnamefont{C.}~\bibnamefont{Dekker}},
  \bibinfo{journal}{Physics Today} \textbf{\bibinfo{volume}{52}},
  \bibinfo{pages}{22} (\bibinfo{year}{1999}).

\bibitem[{\citenamefont{Park et~al.}(2000)\citenamefont{Park, Park, Lim,
  Anderson, Alivisatos, and McEuen}}]{park}
\bibinfo{author}{\bibfnamefont{H.}~\bibnamefont{Park}},
  \bibinfo{author}{\bibfnamefont{J.}~\bibnamefont{Park}},
  \bibinfo{author}{\bibfnamefont{A.~K.~L.} \bibnamefont{Lim}},
  \bibinfo{author}{\bibfnamefont{E.~H.} \bibnamefont{Anderson}},
  \bibinfo{author}{\bibfnamefont{A.~P.} \bibnamefont{Alivisatos}},
  \bibnamefont{and} \bibinfo{author}{\bibfnamefont{P.~L.}
  \bibnamefont{McEuen}}, \bibinfo{journal}{Nature}
  \textbf{\bibinfo{volume}{407}}, \bibinfo{pages}{57} (\bibinfo{year}{2000}).

\bibitem[{\citenamefont{Park et~al.}(2003)\citenamefont{Park, Pasupathy,
  Goldsmith, Soldatov, Chang, Yaish, Sethna, {Abru{\~n}a}, Ralph, and
  McEuen}}]{park2}
\bibinfo{author}{\bibfnamefont{J.}~\bibnamefont{Park}},
  \bibinfo{author}{\bibfnamefont{A.~N.} \bibnamefont{Pasupathy}},
  \bibinfo{author}{\bibfnamefont{J.~I.} \bibnamefont{Goldsmith}},
  \bibinfo{author}{\bibfnamefont{A.~V.} \bibnamefont{Soldatov}},
  \bibinfo{author}{\bibfnamefont{C.}~\bibnamefont{Chang}},
  \bibinfo{author}{\bibfnamefont{Y.}~\bibnamefont{Yaish}},
  \bibinfo{author}{\bibfnamefont{J.~P.} \bibnamefont{Sethna}},
  \bibinfo{author}{\bibfnamefont{H.~D.} \bibnamefont{{Abru{\~n}a}}},
  \bibinfo{author}{\bibfnamefont{D.~C.} \bibnamefont{Ralph}}, \bibnamefont{and}
  \bibinfo{author}{\bibfnamefont{P.~L.} \bibnamefont{McEuen}},
  \bibinfo{journal}{Thin Solid Films} \textbf{\bibinfo{volume}{438-439}},
  \bibinfo{pages}{457} (\bibinfo{year}{2003}).

\bibitem[{\citenamefont{Pasupathy et~al.}(2003)\citenamefont{Pasupathy, Park,
  C.~Chang, Lebedkin, Bialczak, Grose, Donev, Sethna, Ralph, and
  McEuen}}]{pasupathy}
\bibinfo{author}{\bibfnamefont{A.~N.} \bibnamefont{Pasupathy}},
  \bibinfo{author}{\bibfnamefont{J.}~\bibnamefont{Park}},
  \bibinfo{author}{\bibfnamefont{A.~V.~S.} \bibnamefont{C.~Chang}},
  \bibinfo{author}{\bibfnamefont{S.}~\bibnamefont{Lebedkin}},
  \bibinfo{author}{\bibfnamefont{R.~C.} \bibnamefont{Bialczak}},
  \bibinfo{author}{\bibfnamefont{J.~E.} \bibnamefont{Grose}},
  \bibinfo{author}{\bibfnamefont{L.~A.~K.} \bibnamefont{Donev}},
  \bibinfo{author}{\bibfnamefont{J.~P.} \bibnamefont{Sethna}},
  \bibinfo{author}{\bibfnamefont{D.~C.} \bibnamefont{Ralph}}, \bibnamefont{and}
  \bibinfo{author}{\bibfnamefont{P.~L.} \bibnamefont{McEuen}},
  \bibinfo{journal}{cond-mat/0311150}  (\bibinfo{year}{2003}).

\bibitem[{\citenamefont{Yu and Natelson}(2004)}]{natelson}
\bibinfo{author}{\bibfnamefont{L.~H.} \bibnamefont{Yu}} \bibnamefont{and}
  \bibinfo{author}{\bibfnamefont{D.}~\bibnamefont{Natelson}},
  \bibinfo{journal}{Nano Lett.} \textbf{\bibinfo{volume}{4}},
  \bibinfo{pages}{79} (\bibinfo{year}{2004}).

\bibitem[{\citenamefont{Reed and Tour}(2000)}]{reedt}
\bibinfo{author}{\bibfnamefont{M.~A.} \bibnamefont{Reed}} \bibnamefont{and}
  \bibinfo{author}{\bibfnamefont{J.~M.} \bibnamefont{Tour}},
  \bibinfo{journal}{Scientific American} \textbf{\bibinfo{volume}{282}},
  \bibinfo{pages}{86} (\bibinfo{year}{2000}).

\bibitem[{\citenamefont{Hettler et~al.}(2002)\citenamefont{Hettler, Schoeller,
  and Wenzel}}]{schoeller1}
\bibinfo{author}{\bibfnamefont{M.~H.} \bibnamefont{Hettler}},
  \bibinfo{author}{\bibfnamefont{H.}~\bibnamefont{Schoeller}},
  \bibnamefont{and} \bibinfo{author}{\bibfnamefont{W.}~\bibnamefont{Wenzel}},
  \bibinfo{journal}{Europhys. Lett.} \textbf{\bibinfo{volume}{57}},
  \bibinfo{pages}{571} (\bibinfo{year}{2002}).

\bibitem[{\citenamefont{Hettler et~al.}(2003)\citenamefont{Hettler, Wenzel,
  Wegewijs, and Schoeller}}]{schoeller2}
\bibinfo{author}{\bibfnamefont{M.~H.} \bibnamefont{Hettler}},
  \bibinfo{author}{\bibfnamefont{W.}~\bibnamefont{Wenzel}},
  \bibinfo{author}{\bibfnamefont{M.~R.} \bibnamefont{Wegewijs}},
  \bibnamefont{and}
  \bibinfo{author}{\bibfnamefont{H.}~\bibnamefont{Schoeller}},
  \bibinfo{journal}{Phys. Rev. Lett.} \textbf{\bibinfo{volume}{90}},
  \bibinfo{pages}{076805} (\bibinfo{year}{2003}).

\bibitem[{\citenamefont{Boese and Schoeller}(2001)}]{schoeller3}
\bibinfo{author}{\bibfnamefont{D.}~\bibnamefont{Boese}} \bibnamefont{and}
  \bibinfo{author}{\bibfnamefont{H.}~\bibnamefont{Schoeller}},
  \bibinfo{journal}{Europhys. Lett.} \textbf{\bibinfo{volume}{54}},
  \bibinfo{pages}{668} (\bibinfo{year}{2001}).

\bibitem[{\citenamefont{Mitra et~al.}(2004)\citenamefont{Mitra, Aleiner, and
  Millis}}]{aleiner}
\bibinfo{author}{\bibfnamefont{A.}~\bibnamefont{Mitra}},
  \bibinfo{author}{\bibfnamefont{I.}~\bibnamefont{Aleiner}}, \bibnamefont{and}
  \bibinfo{author}{\bibfnamefont{A.~J.} \bibnamefont{Millis}},
  \bibinfo{journal}{Phys. Rev. B} \textbf{\bibinfo{volume}{69}},
  \bibinfo{pages}{245302} (\bibinfo{year}{2004}).

\bibitem[{\citenamefont{Aji et~al.}(2003)\citenamefont{Aji, Moore, and
  Varma}}]{varma}
\bibinfo{author}{\bibfnamefont{V.}~\bibnamefont{Aji}},
  \bibinfo{author}{\bibfnamefont{J.~E.} \bibnamefont{Moore}}, \bibnamefont{and}
  \bibinfo{author}{\bibfnamefont{C.~M.} \bibnamefont{Varma}},
  \bibinfo{journal}{cond-mat/0302222}  (\bibinfo{year}{2003}).

\bibitem[{\citenamefont{Braig and Flensberg}(2003)}]{flensb1}
\bibinfo{author}{\bibfnamefont{S.}~\bibnamefont{Braig}} \bibnamefont{and}
  \bibinfo{author}{\bibfnamefont{K.}~\bibnamefont{Flensberg}},
  \bibinfo{journal}{Phys. Rev. B} \textbf{\bibinfo{volume}{68}},
  \bibinfo{pages}{205324} (\bibinfo{year}{2003}).

\bibitem[{\citenamefont{Flensberg}(2003)}]{flensb2}
\bibinfo{author}{\bibfnamefont{K.}~\bibnamefont{Flensberg}},
  \bibinfo{journal}{Phys. Rev. B} \textbf{\bibinfo{volume}{68}},
  \bibinfo{pages}{205323} (\bibinfo{year}{2003}).

\bibitem[{\citenamefont{Xue and Ratner}(2003)}]{ratner2}
\bibinfo{author}{\bibfnamefont{Y.}~\bibnamefont{Xue}} \bibnamefont{and}
  \bibinfo{author}{\bibfnamefont{M.~A.} \bibnamefont{Ratner}},
  \bibinfo{journal}{Phys. Rev. B} \textbf{\bibinfo{volume}{68}},
  \bibinfo{pages}{115407} (\bibinfo{year}{2003}).

\bibitem[{\citenamefont{{Di Ventra} et~al.}(2002)\citenamefont{{Di Ventra},
  Lang, and Pantelides}}]{diventra}
\bibinfo{author}{\bibfnamefont{M.}~\bibnamefont{{Di Ventra}}},
  \bibinfo{author}{\bibfnamefont{N.~D.} \bibnamefont{Lang}}, \bibnamefont{and}
  \bibinfo{author}{\bibfnamefont{S.~T.} \bibnamefont{Pantelides}},
  \bibinfo{journal}{Chem. Phys.} \textbf{\bibinfo{volume}{281}},
  \bibinfo{pages}{189} (\bibinfo{year}{2002}).

\bibitem[{\citenamefont{Evers et~al.}(2003)\citenamefont{Evers, Weigend, and
  Koentopp}}]{evers}
\bibinfo{author}{\bibfnamefont{F.}~\bibnamefont{Evers}},
  \bibinfo{author}{\bibfnamefont{F.}~\bibnamefont{Weigend}}, \bibnamefont{and}
  \bibinfo{author}{\bibfnamefont{M.}~\bibnamefont{Koentopp}},
  \bibinfo{journal}{cond-mat/0312122}  (\bibinfo{year}{2003}).

\bibitem[{\citenamefont{Beenakker and Staring}(1992)}]{beenakker1}
\bibinfo{author}{\bibfnamefont{C.~W.~J.} \bibnamefont{Beenakker}}
  \bibnamefont{and} \bibinfo{author}{\bibfnamefont{A.~A.~M.}
  \bibnamefont{Staring}}, \bibinfo{journal}{Phys. Rev. B}
  \textbf{\bibinfo{volume}{46}}, \bibinfo{pages}{9667} (\bibinfo{year}{1992}).

\bibitem[{\citenamefont{Staring et~al.}(1993)\citenamefont{Staring, Molenkamp,
  Alphenaar, {van Houten}, Buyk, Mabesoone, Beenakker, and Foxon}}]{beenakker2}
\bibinfo{author}{\bibfnamefont{A.~A.~M.} \bibnamefont{Staring}},
  \bibinfo{author}{\bibfnamefont{L.~W.} \bibnamefont{Molenkamp}},
  \bibinfo{author}{\bibfnamefont{B.~W.} \bibnamefont{Alphenaar}},
  \bibinfo{author}{\bibfnamefont{H.}~\bibnamefont{{van Houten}}},
  \bibinfo{author}{\bibfnamefont{O.~J.~A.} \bibnamefont{Buyk}},
  \bibinfo{author}{\bibfnamefont{M.~A.~A.} \bibnamefont{Mabesoone}},
  \bibinfo{author}{\bibfnamefont{C.~W.~J.} \bibnamefont{Beenakker}},
  \bibnamefont{and} \bibinfo{author}{\bibfnamefont{C.~T.} \bibnamefont{Foxon}},
  \bibinfo{journal}{Europhys. Lett.} \textbf{\bibinfo{volume}{22}},
  \bibinfo{pages}{57} (\bibinfo{year}{1993}).

\bibitem[{\citenamefont{Averin and Nazarov}(1990)}]{averin2}
\bibinfo{author}{\bibfnamefont{D.~V.} \bibnamefont{Averin}} \bibnamefont{and}
  \bibinfo{author}{\bibfnamefont{Y.~V.} \bibnamefont{Nazarov}},
  \bibinfo{journal}{Phys. Rev. Lett.} \textbf{\bibinfo{volume}{65}},
  \bibinfo{pages}{2446} (\bibinfo{year}{1990}).

\bibitem[{\citenamefont{Turek and Matveev}(2002)}]{turek}
\bibinfo{author}{\bibfnamefont{M.}~\bibnamefont{Turek}} \bibnamefont{and}
  \bibinfo{author}{\bibfnamefont{K.~A.} \bibnamefont{Matveev}},
  \bibinfo{journal}{Phys. Rev. B} \textbf{\bibinfo{volume}{65}},
  \bibinfo{pages}{115332} (\bibinfo{year}{2002}).

\bibitem[{\citenamefont{Matveev and Andreev}(2002)}]{andreev}
\bibinfo{author}{\bibfnamefont{K.~A.} \bibnamefont{Matveev}} \bibnamefont{and}
  \bibinfo{author}{\bibfnamefont{A.~V.} \bibnamefont{Andreev}},
  \bibinfo{journal}{Phys. Rev. B} \textbf{\bibinfo{volume}{66}},
  \bibinfo{pages}{045301} (\bibinfo{year}{2002}).

\bibitem[{foo()}]{footnote}
\bibinfo{howpublished}{At low temperatures the cotunneling terms increase due
  to higher order contributions that lead to strong Kondo correlations near the
  Kondo temperature $T_K$. In this article, we assume temperatures high
  compared to the Kondo temperature, $T \gg T_K$. The thermopower for $T\agt
  T_K$ can be estimated by substituting the cotunneling rate by an effective
  one, which is renormalized by the higher-order terms, cf., e.g., M. Pustilnik
  and L. I. Glazman, J. Phys. Condens. Matter \textbf{16}, R513 (2004).}

\bibitem[{\citenamefont{Mahan}(1990)}]{mahan}
\bibinfo{author}{\bibfnamefont{G.~D.} \bibnamefont{Mahan}},
  \emph{\bibinfo{title}{Many-Particle Physics}} (\bibinfo{publisher}{Plenum
  Press, New York}, \bibinfo{year}{1990}), chap. \bibinfo{chapter}{4.3}.

\bibitem[{\citenamefont{Averin}(1994)}]{averin}
\bibinfo{author}{\bibfnamefont{D.~V.} \bibnamefont{Averin}},
  \bibinfo{journal}{Physica B} \textbf{\bibinfo{volume}{194-196}},
  \bibinfo{pages}{979} (\bibinfo{year}{1994}).

\bibitem[{\citenamefont{Blum}(1981)}]{blum}
\bibinfo{author}{\bibfnamefont{K.}~\bibnamefont{Blum}},
  \emph{\bibinfo{title}{Density Matrix Theory and Applications}}
  (\bibinfo{publisher}{Plenum Press, New York}, \bibinfo{year}{1981}),
  chap.~\bibinfo{chapter}{7}.

\bibitem[{\citenamefont{Abramowitz and {Stegun (eds.)}}(1972)}]{abramo}
\bibinfo{author}{\bibfnamefont{M.}~\bibnamefont{Abramowitz}} \bibnamefont{and}
  \bibinfo{author}{\bibfnamefont{I.~A.} \bibnamefont{{Stegun (eds.)}}},
  \emph{\bibinfo{title}{Handbook of Mathematical Functions}}
  (\bibinfo{publisher}{Dover Publications, New York}, \bibinfo{year}{1972}),
  chap. \bibinfo{chapter}{6.4}, p. \bibinfo{pages}{260}.

\end{thebibliography}
\end{document}